\documentclass[11pt]{article}

\newcommand{\be}{\begin{equation}}
\newcommand{\ee}{\end{equation}}
\newcommand{\bea}{\begin{eqnarray}}
\newcommand{\eea}{\end{eqnarray}}
\newcommand{\ba}{\begin{array}}
\newcommand{\ea}{\end{array}}
\newcommand{\hone}{h_{1,1}}
\newcommand{\htwo}{h_{2,1}}
\newcommand{\M}{\mathcal{M}}
\newcommand{\N}{\mathcal{N}}

\newcommand{\F}{\mathcal{F}}
\newcommand{\A}{\mathcal{A}}

\long\def\symbolfootnote[#1]#2{\begingroup%
\def\thefootnote{\fnsymbol{footnote}}\footnote[#1]{#2}\endgroup}

\begin{document}

\thispagestyle{empty} \vspace{40pt} \hfill{hep-th/0701060}

\vspace{128pt}

\begin{center}
\textbf{\Large Five dimensional 2-branes and \\ \vspace{10pt} the universal hypermultiplet }\\

\vspace{40pt}

Moataz H. Emam\symbolfootnote[1]{Electronic address: {\tt moataz.emam@cortland.edu}}

\vspace{12pt} \textit{Department of Physics}\\
\textit{SUNY College at Cortland}\\
\textit{Cortland, NY 13045, USA}\\
\end{center}

\vspace{40pt}

\begin{abstract}

We present a discussion of black 2-branes coupled to the fields of
the universal hypermultiplet of ungauged $\N=2$ supergravity
theory in five dimensions. Using a general ansatz dependent on a
spherically symmetric harmonic function, we show that there are
exactly two such solutions, both of which can be thought
of as arising from the dimensional reduction of either M2- or
M5-branes over special Lagrangian cycles of a Calabi-Yau 3-fold, confirming previous results.
By relaxing some of the constraints on the ansatz, we proceed to
find a more general solution carrying both M-brane charges and
discuss its properties, as well as its relationship to Euclidean instantons.

\end{abstract}

\newpage


\tableofcontents

\vspace{15pt}

\section{Introduction}

The search for nonperturbative solutions of string/supergravity
theories is an open problem in theoretical physics with various
important applications. For example interest in $\N=2$ branes, our
focus in this paper, soared in recent years because of their
relevance to the conjectured equivalence between string theory on
anti-de Sitter space and certain superconformal gauge theories
living on the boundary of the space (the AdS/CFT duality)
\cite{Maldacena:2001uc}. Some solitonic solutions in $D=4,5$ also
allow for interpretation as the dimensional reduction of branes
wrapping supersymmetric cycles of manifolds with restricted
holonomy. For example, M-branes wrapping K\"{a}hler cycles of a
Calabi-Yau 3-fold $\M$ \cite{Cho:2000hg} dimensionally reduce to
black holes and strings coupled to the vector multiplets of five
dimensional $\N=2$ supergravity \cite{Kastor:2003jy}, while
M-branes wrapping special Lagrangian cycles reduce to
configurations carrying charge under the hypermultiplet scalars
\cite{Martelli:2003ki,Fayyazuddin:2005as,Emam:2005bh,Emam:2006sr}.
Studying how higher dimensional results are related to lower
dimensional ones may eventually provide clues to the explicit
structure of the compact space and the choice of compactification
mechanism, thereby contributing to more understanding of the
string theory landscape.

An abundance of work on gravitating solitonic solutions of $\N=2$
supergravity (SUGRA) in four and five dimensions exist in the
literature. One notes, however, that most of these study couplings
with the vector multiplets sector (for example, see
\cite{Behrndt:1997fq,Sabra:1997kq,Sabra:1997dh,Behrndt:1997ny,Sabra:1997yd,Behrndt:1997as,Behrndt:1998eq,Behrndt:1998ns}).
In contrast, hypermultiplets-coupled solutions are quite rare
(\textit{e.g.}
\cite{Behrndt:2000km,Ceresole:2001wi,Behrndt:2002ee,Cacciatori:2002qx}).
As such, this paper is one in a series intended to filling this gap
as well as generalizing previous results \cite{Emam:2005bh}.

The relative rarity of hypermultiplets-related work may be
ascribed to the fact that the scalar fields of the hypermultiplets
sector parameterize a quaternionic manifold, which is generally
difficult to deal with. Luckily there exists a duality
transformation, known as the c-map, which relates the $D=5$
hypermultiplets' quaternionic structure to the more well
understood $D=4$ special K\"{a}hler structure parameterized by the
four dimensional vector multiplets (\textit{e.g.}
\cite{DeJaegher:1997ka,deWit:1995tf}). This means that one can use
the techniques of special geometry, usually reserved for the
vector multiplets, to study the hypermultiplets and classify their
brane-coupled behavior. This was first performed in
\cite{Gutperle:2000sb,Gutperle:2000ve}, where instanton solutions
of Euclidean $\N=2$ $D=5$ supergravity were found using special
geometry methods. The 2-branes we have studied in
\cite{Emam:2005bh,Emam:2006sr}, as well as the new result in this
paper, are magnetically dual to these instantons. The general
topic of $\N=2$ instantons was studied in various sources (see
\cite{Gutperle:2000sb} and the references within).

In \cite{Emam:2005bh} we found two solutions representing the
coupling of 2-branes to the $D=5$ universal hypermultiplet fields
by dimensionally reducing particular M-brane configurations (which
were constructed using the calibrations technique). In this paper,
we re-derive these solutions using an ansatz based on a radial
function satisfying the Laplace equation in the brane's transverse
space. We demonstrate that these solutions are the only possible
ones satisfying the ansatz and are characterized by the vanishing
of either the M2- or M5-brane charges (inherited from $D=11$ via
dimensional reduction). We then proceed to generalize to a new
solution carrying both charges and discuss some of its properties.
We also note that changing certain choices leads to the dual
Euclidean space solutions, \emph{i.e.} instantons. This
straightforwardly follows from the (Poincar\'{e})$_3 \times
SO\left( 2 \right)$ invariance of the metric.

\section{Five dimensional $\N=2$ supergravity}

The form of the ungauged $\N=2$ supergravity theory living in five
dimensions may be understood in geometric terms by considering it
as the dimensional reduction of $D=11$ supergravity over
supersymmetric cycles of a Calabi-Yau 3-fold $\M$. The resulting
theory exhibits a surprisingly rich structure that follows
directly from the intricate topology of the compact space. The
matter content of the theory is composed of $(\hone-1)$ vector
multiplets and $(\htwo+1)$ hypermultiplets
\cite{Cadavid:1995bk,Aspinwall:2000fd,Ferrara:1996hh,Lukas:1998tt,Papadopoulos:1995da};
the $h$'s being the Hodge numbers of $\M$. These two sectors
decouple such that one may consistently set either one of them to zero.
Furthermore, compactification over a rigid Calabi-Yau with
constant complex structure moduli ($\htwo=0$) excites only the
so-called universal hypermultiplet (UH), which is the only one we
keep in our presentation. Throughout, we will use
$\varepsilon_{M_1 M_2  \cdots M_D }$ to denote the totally
antisymmetric Levi-Civita symbol such that $\varepsilon _{012
\cdots }  = + 1$.

In what follows, we largely use the formulation of
\cite{Gutperle:2000sb}. The bosonic fields of $D=11$ supergravity
theory are the metric and a $3$-form gauge potential
$\mathcal{A}$. The action is given by
\be\label{eleven}
    S_{11}  = \int_{11} \left( { {\mathcal{R}\star 1 -
    \frac{1}{2}\F \wedge \star \F -
    \frac{1}{6}\A \wedge \F \wedge
    \F}}\right),
\ee
where $\star$ is the Hodge duality operator. The $D=5$ couplings
of gravity to the universal hypermultiplet follow from eleven
dimensional fields of the form
\bea\label{universal-metric}
    ds^2_{11} & = &e^{\frac{2}{3}\sigma}g_{MN}dx^M dx^N + e^{ - \frac{\sigma }{3}}ds_{CY}^2,\quad\quad M,N=0,\ldots,4 \nonumber\\
    \mathcal{A} & = & {1\over 3!}A_{MNP} dx^M\wedge dx^N\wedge dx^P +
    {1\over \sqrt{2} }\left(\chi\Omega + \bar\chi\bar\Omega\right    )\nonumber\\
    \mathcal{F} & = & d\mathcal{A}={1\over 4!}F_{LMNP} dx^L\wedge dx^M\wedge dx^N\wedge dx^P +
    {1\over \sqrt{2} }\left(d\chi\wedge\Omega + d\bar\chi\wedge\bar\Omega\right ),
\eea
where $ds_{CY}^2$ is a fixed Ricci flat metric on the Calabi-Yau
space $\M$ and $\Omega$ is the unique holomorphic $3$-form on
$\M$. The $D=5$ fields of the universal hypermultiplet are the
real overall volume modulus $\sigma$ of the Calabi-Yau space, the
$D=5$ 3-form gauge potential $A$ and the pseudo-scalar
axions $(\chi, \bar\chi)$, assumed conjugate to each other. The dual to $A$ is a scalar field known
as the universal axion $a$. The UH scalars $(\sigma, a, \chi,
\bar\chi)$ parameterize the quaternionic manifold $SU(2,1)/U(2)$
\cite{Ferrara:1989ik, Cecotti:1988qn}. The $D=5$ action is then
found to be (dimensional reduction detailed in
\cite{Emam:2004nc}):
\be
    S_5  = \int\limits_5 {\left[ {R\star 1 - \frac{1}{2}d\sigma  \wedge \star d\sigma  - \frac{1}{2}e^{ - 2\sigma } F \wedge \star F - e^\sigma  d\chi  \wedge \star d\bar \chi }
    - \frac{i}{2}F \wedge \left( {\chi d\bar \chi  - \bar \chi d\chi } \right)\right]}.
\ee

The equations of motion of $\sigma$, $A$ and $(\chi, \bar \chi)$ are,
respectively:
\bea
    \left( {\Delta \sigma } \right)\star 1 - e^\sigma  d\chi  \wedge \star d \bar \chi  + e^{ - 2\sigma } F \wedge \star F &=& 0 \label{sigmaeom}\\
    d^{\dag} \left[ {e^{ - 2\sigma } F + \frac{i}{2}\star\left( {\chi d\bar \chi  - \bar \chi d\chi } \right)} \right] &=& 0 \label{Feom}\\
    d^{\dag} \left[ {e^\sigma  d\chi  + i\chi \star F} \right] &=& 0 \nonumber\\
    d^{\dag} \left[ {e^\sigma  d\bar \chi  - i\bar \chi \star F} \right] &=& 0, \label{chieom2}
\eea
where $d^\dag$ is the adjoint exterior derivative, $\Delta$ is the
Laplace-De Rahm operator and the Bianchi identity $dF=0$ was used
to simplify (\ref{chieom2}). The form of equations (\ref{Feom})
and (\ref{chieom2}) is suggestive. It implies that the quantities
inside the derivatives are conserved currents:
\bea
 \mathcal{J}_2  &=& e^{ - 2\sigma } F + \frac{i}{2}\star\left( {\chi   d\bar \chi  - \bar \chi   d\chi } \right) \nonumber\\
 \mathcal{J}_5 &=& e^\sigma   d\chi  + i\chi\star  F, \quad\quad\quad
 \bar \mathcal{J}_5 = e^\sigma   d\bar \chi  - i\bar \chi\star
 F.\label{Noethercurrents}
\eea

These are, in fact, conserved Noether currents. The charges
associated with them are found in the usual way by:
\bea
 \mathcal{Q}_2  &=& \int {\mathcal{J}_2 },\nonumber\\
 \mathcal{Q}_5  &=& \int { \mathcal{J}_5 } ,\quad \quad \quad \bar \mathcal{Q}_5  = \int {\bar \mathcal{J}_5
 }.\label{Charges}
\eea

These charges (more accurately charge \textit{densities})
correspond to certain isometries of the quaternionic manifold
$SU(2,1)/U(2)$ as discussed in various sources
\cite{Ferrara:1989ik, Cecotti:1988qn}. From a five dimensional
perspective, they can be thought of as the result of the
invariance of the action under certain infinitesimal shifts of $A$
and $\left(\chi, \bar\chi\right)$
\cite{Gutperle:2000sb,Gutperle:2000ve}. The geometric way of
understanding these charges is in noting that they descend from
the eleven dimensional electric and magnetic M-brane charges,
hence the $\left(2,5\right)$ labels\footnote{This is the reverse
situation to that of \cite{Gutperle:2000sb}, where the (dual)
Euclidean theory was studied.}. M2-branes wrapping special
Lagrangian cycles of $\M$ generate $\mathcal{Q}_2$ while the
wrapping of M5-branes results in non-vanishing
$\left(\mathcal{Q}_5, \bar \mathcal{Q}_5\right)$.

Finally, the full action is invariant under the following set of
supersymmetry (SUSY) variations:
\bea
    \delta _\epsilon  \psi ^1  &=& D\epsilon _1  + \frac{i}{4}e^{ - \sigma } \star F\epsilon _1  - \frac{{e^{\frac{\sigma }{2}} }}{{\sqrt 2 }}d\chi \epsilon _2  \nonumber\\
    \delta _\epsilon  \psi ^2  &=& D\epsilon _2  - \frac{i}{4}e^{ - \sigma } \star F\epsilon _2  + \frac{{e^{\frac{\sigma }{2}} }}{{\sqrt 2 }}d\bar \chi \epsilon _1 \label{gravitinisusy} \\
    \delta _\epsilon  \xi _1  &=& \frac{1}{2}\left[ {\left( {\partial _M \sigma } \right) - ie^{ - \sigma } \left( {\star F} \right)_M } \right]\Gamma ^M \epsilon _1  + \frac{{e^{\frac{\sigma }{2}} }}{{\sqrt 2 }}\left( {\partial _M \chi } \right)\Gamma ^M \epsilon _2  \nonumber\\
    \delta _\epsilon  \xi _2  &=& \frac{1}{2}\left[ {\left( {\partial _M \sigma } \right) + ie^{ - \sigma } \left( {\star F} \right)_M } \right]\Gamma ^M \epsilon _2  - \frac{{e^{\frac{\sigma }{2}} }}{{\sqrt 2 }}\left( {\partial _M \bar \chi } \right)\Gamma ^M \epsilon _1, \label{hyperSUSY}
\eea
where
\be
    D = dx^M \left( {\partial _M  + \frac{1}{4}\omega _M^{\,\,\,\,\,\hat M\hat N} \Gamma _{\hat M\hat N} }
    \right)
\ee
as usual, $\psi$ and $\xi$ are the gravitini and hyperini fermions
respectively, the $\epsilon$'s are the $\N=2$ SUSY spinors,
$\omega$ is the spin connection and the hatted indices are frame
indices in a flat tangent space.

\section{\label{HarmonicAnsatzMain}The main ansatz and other preliminaries}

The most general form of a metric satisfying (Poincar\'{e})$_3
\times SO\left( 2 \right)$ invariance in $D=5$ is:
\begin{equation}\label{metricansatzoriginal}
    ds^2  = e^{2C\sigma \left( r \right) } \eta _{ab} dx^a dx^b  + e^{2B\sigma \left( r \right) } \delta _{\mu \nu } dx^\mu  dx^\nu  ,\,\,\,\,\,\,\,\,\,\,\,\,\,\,a,b = 0,1,2\,\,\,\,\,\,\,\,\,\,\mu ,\nu  =
    3,4
\end{equation}
where $C $ and $B$ are constants, $(x^1,x^2)$ define the
directions tangent to the brane and $(x^3,x^4)$ those transverse
to it. From the point of view of the $(x^3,x^4)$ plane, the brane
appears as a point and one may construct ordinary polar
coordinates $\left( r, \theta\right)$ in the plane with the brane
at the origin, such that $r = \left( {\delta ^{\mu \nu } x_\mu
x_\nu  } \right)^{{1 \mathord{\left/
 {\vphantom {1 2}} \right.
 \kern-\nulldelimiterspace} 2}} $ and $\tan \theta  = {{x_4 } \mathord{\left/
 {\vphantom {{x_4 } {x_3 }}} \right. \kern-\nulldelimiterspace} {x_3 }}$. It turns out that the constant $C$ is constrained to vanish by both
the Einstein equations and the SUSY condition\footnote{This is no
surprise, since $\delta \psi=0$ automatically satisfies
$G_{MN}=8\pi T_{MN}$ \cite{Kaya:1999mm}.} $\delta \psi=0$. We then
set $C=0$ from the start:
\begin{equation}\label{metricansatz}
    ds^2  = \eta _{ab} dx^a dx^b  + e^{2B\sigma \left( r \right) } \delta _{\mu \nu } dx^\mu  dx^\nu.
\end{equation}

We also adopt the following form for $F$ throughout:
\be
     F  = \frac{\kappa }{{3!}} \varepsilon _{abc} f_\mu ^{\,\,\,\,\nu } \left( {\partial _\nu  H^m } \right) dx^\mu\wedge  dx^a\wedge dx^b\wedge dx^c, \label{ansatze1}
\ee
where $m$ and $\kappa$ are real constants to be determined, $f_\mu
^{\,\,\,\,\nu }$ is either $\delta_\mu^\nu$ or
$\varepsilon_\mu^{\;\;\;\nu}$ and the radial function $H(r)$ is
assumed harmonic in the transverse directions; \emph{i.e.}
\bea
    \Delta H &=& \delta ^{\mu \nu } \left( {\partial _\mu  \partial _\nu  H}
    \right)\nonumber\\
    &=&\frac{1}{r}\frac{{d \,\,}}{{d r}}\left( {r\frac{{d H}}{{d r}}} \right) =    0\nonumber\\
    \to\quad H\left(r\right ) &=& 1 + q\ln r.\label{harmonic}
\eea

Here $q$ is related to the charges of the dimensionally reduced
M-branes and we have chosen $H\left(1\right)=1$ for simplicity. We
note as well that it is possible to generalize this ansatz to a
multi-centered solution simply by redefining:
\be
    H\left( {{\bf x} - {\bf x}_i } \right) = 1 + \sum\limits_{i = 1}^N {q_i \ln \left| {{\bf x} - {\bf x}_i } \right|},\label{multicenter} \ee
where $\bf x$ is the position vector in the $(x^3,x^4)$ plane and
${\bf x}_i$ are the position vectors of a number $N$ of 2-branes,
each carrying a charge $q_i$. The presence of harmonic functions
is well understood in the literature as characteristic of
supersymmetric Bogomol'nyi-Prasad-Sommerfield (BPS) solutions,
making the harmonic condition a desirable property. They also
indicate that the multi-centered version of the solution satisfies
a Newtonian `no-force' condition (an example in the case of five
dimensional black holes is \cite{Liu:2000ah}). In this work,
however, we will restrict ourselves to studying single-centered
solutions.

Demanding that (\ref{ansatze1}) satisfies the Bianchi identity
$dF=0$ leads to:
\be
    \kappa\left( {m - 1} \right)\varepsilon ^{\mu\alpha } f_\alpha
    ^{\,\,\,\,\nu } \left( {\partial _\mu  \ln H} \right)\left(
    {\partial _\nu  \ln H} \right) = 0,
\ee
which, for nonvanishing $\kappa$, is identically satisfied for the
choice $f_\mu ^{\,\,\,\,\nu } = \delta _\mu ^\nu  $ but requires
$m=1$ for $f_\mu ^{\,\,\,\,\nu } = \varepsilon _\mu ^{\,\,\,\,\nu
} $. Now, consider the field equation (\ref{Feom}). The second
term vanishes as a consequence of the Bianchi identities on
$\left(\chi, \bar\chi\right)$ as well as
\be
    d\chi  \wedge d\bar \chi  = 0,\label{dchiwedgedchibar}
\ee
while the first term leads to
\bea
    d\star F - 2d\sigma  \wedge\star F &=& 0\nonumber\\
    \kappa f^{\mu \nu } x_\mu  x_\nu  \left[ {\left( {m - 1} \right)\frac{q}{{rH}}- 2\frac{{d\sigma }}{{dr}} } \right]
    &=&    0,\label{diffeqsigma1}
\eea
where the assumption for a radial dilaton (\ref{metricansatz}) was
used. Once again assuming nonvanishing $\kappa$
(\ref{diffeqsigma1}) is identically satisfied for the choice
$f_\mu ^{\,\,\,\,\nu } = \varepsilon _\mu ^{\,\,\,\,\nu } $.
However for $f_\mu ^{\,\,\,\,\nu } = \delta _\mu ^\nu $ it yields
a first order differential equation for $\sigma$ with the
solution:
\be
    \sigma \left( r \right) = \frac{1}{2}\left( {m - 1} \right)\ln H,\label{sigmafordelta} \ee
where we have chosen $\sigma\left(1\right)=0$ for simplicity. We
then generally assume the following form for the dilaton
\be
    \sigma \left( r \right) = n\ln H,\label{sigmageneral}
\ee
where $n$ is a constant that is equal to $\frac{1}{2}\left( {m -
1} \right)$ only for $\kappa\ne 0$ and $f_\mu ^{\,\,\,\,\nu } =
\delta _\mu ^\nu $.

The solutions we seek are black 2-branes satisfying the BPS
condition, breaking half of the supersymmetries of the theory.
This is guaranteed by the vanishing of the variation of gravitini
and hyperini backgrounds, \textit{i.e.} $\delta \psi=0$ and
$\delta \xi=0$. In addition, we will use the remaining UH field
equations and Einstein's equations $G_{MN}=8\pi T_{MN}$ to find
constraints on the unknown constants in the ansatz as well as the
form of the axions $\left(\chi, \bar\chi\right)$. The bits and
pieces we will need are as follows: The f\"{u}nfbeins and
Christoffel symbols are:
\begin{eqnarray}
    e_{\;\;b}^{\hat a}  &=& \eta _b^{\hat a},\quad \quad \quad \quad
    e_{\;\;\nu}    ^{\hat     \mu }  = e^{B\sigma } \delta _\nu ^{\hat \mu }  \nonumber \\
    \Gamma _{\nu \rho }^\mu   &=& B\left[ {\delta _\nu ^\mu  \left( {\partial _\rho  \sigma
    }     \right) + \delta _\rho ^\mu  \left( {\partial _\nu  \sigma } \right) - \delta
    _{\nu \rho } \delta ^{\mu \alpha } \left( {\partial _\alpha  \sigma } \right)}
    \right],
\end{eqnarray}
resulting in the following spin connections and covariant
derivatives:
\begin{eqnarray}
    \omega _\alpha ^{\;\;\hat \beta \hat \gamma }&=& B\left( {\delta _\alpha ^{\hat \beta }
    \delta ^{\hat \gamma \rho }  - \delta ^{\hat \beta \rho } \delta _\alpha ^{\hat \gamma
    }     } \right)\left( {\partial _\rho  \sigma } \right)\nonumber \\
     D_a   &=& \partial _a  \nonumber \\
    D_\mu   &=& \partial _\mu   + \frac{B}{2}\left( {\partial _\nu  \sigma
    }     \right){\Gamma _\mu } ^\nu,
\end{eqnarray}
as well as the Dirac matrices projection conditions\footnote{The
Einstein summation convention is \emph{not} used over the index
$s$.}:
\begin{eqnarray}
    \Gamma _{\hat \mu \hat \nu } \epsilon _s  &= b_s \varepsilon _{\hat
    \mu     \hat \nu } \epsilon _s,  \quad\quad\quad\quad s &=(1,2),\quad b_s=\pm
    i \nonumber \\
    \Gamma _\mu^{\;\;\;\nu} \epsilon _s  &= b_s {\varepsilon_\mu}^{\;\nu}\epsilon
    _s, \quad\quad
    \Gamma ^\mu  \epsilon _s  &=  - b_s {\varepsilon_\nu}^{\;\mu} \Gamma
    ^\nu          \epsilon _s.\label{projection}
\end{eqnarray}

Finally, the Einstein equations have the following components:
\bea
 G_{ab}  &=& B\eta _{ab} g^{\mu \nu } \left( {\partial _\mu  \partial _\nu  \sigma } \right),\quad \quad \quad \quad G_{\mu \nu }  = 0 \nonumber\\
 8\pi T_{ab}  &=&  - \frac{1}{4}\eta _{ab} \left( {\partial ^\mu  \sigma } \right)\left( {\partial _\mu  \sigma } \right) + \frac{1}{4}e^{ - 2\sigma } F_{\mu dca} F_{\,\,\,\,\,\,\,\,\,\,b}^{\mu dc}  - \frac{1}{{24}}e^{ - 2\sigma } \eta _{ab} F_{\mu dce} F^{\mu dce}  - \frac{1}{2}e^\sigma  \eta _{ab} \left( {\partial ^\mu  \chi } \right)\left( {\partial _\mu  \bar \chi } \right) \nonumber\\
 8\pi T_{\mu \nu }  &=& \frac{1}{2}\left( {\partial _\mu  \sigma } \right)\left( {\partial _\nu  \sigma } \right) - \frac{1}{4}g_{\mu \nu } \left( {\partial ^\alpha  \sigma } \right)\left( {\partial _\alpha  \sigma } \right) + \frac{1}{{12}}e^{ - 2\sigma } F_{\mu abc} F_\nu ^{\,\,\,\,abc}  - \frac{1}{{24}}e^{ - 2\sigma } g_{\mu \nu } F_{\alpha abc} F^{\alpha abc}  \nonumber\\
  & &+ e^\sigma  \left( {\partial _\mu  \chi } \right)\left( {\partial _\nu  \bar \chi } \right) - \frac{1}{2}e^\sigma  g_{\mu \nu } \left( {\partial ^\alpha  \chi } \right)\left( {\partial _\alpha  \bar \chi }
  \right).
\eea

\section{An ansatz for the axions}\label{Derivation}

In this section we assume the form of the axion fields
$\left( {\chi ,\bar \chi } \right)$ to be:
\be
    d\chi = \omega l_\mu ^{\,\,\,\,\nu} \left( {\partial _\nu  H^p } \right)dx^\mu,\quad \quad\quad d{\bar \chi} = \bar\omega l_\mu
    ^{\,\,\,\,\nu } \left( {\partial _\nu  H^p } \right)dx^\mu,\label{ansatze2}
\ee
where $p$ and the complex $\omega$ are constants to be determined and $l_\mu
^{\,\,\,\,\nu }$ is either $\delta_\mu^\nu$ or
$\varepsilon_\mu^{\;\;\;\nu}$. We will derive all the possible solutions that satisfy this assumption. This will include both 2-brane solutions as well as the dual instanton solutions. It is straightforward to check that
(\ref{ansatze2}) trivially satisfies (\ref{dchiwedgedchibar}). We
now systematically go through the remaining field equations,
inserting the ans\"{a}tze (\ref{ansatze1}, \ref{sigmageneral},
\ref{ansatze2}) and keeping track of the cases that arise due to
the specific choices of $f_\mu ^{\,\,\,\,\nu }$ and $l_\mu
^{\,\,\,\,\nu }$. The Bianchi identity $d^2\chi=0$ gives
\be
    \omega \left( {p - 1} \right)\varepsilon ^{\mu \alpha } l_\alpha ^{\,\,\,\,\nu } \left( {\partial _\mu  \ln H} \right)\left( {\partial _\nu  \ln H} \right) =
    0,
\ee
which is satisfied identically if $l_\mu ^{\,\,\,\,\nu
}=\delta_\mu ^\nu$ but requires
\be
    \omega \left( {p - 1} \right)=0
\ee
for $l_\mu^{\,\,\,\,\nu }=\varepsilon_\mu ^{\,\,\,\,\nu }$. The
$\sigma$ equation of motion (\ref{sigmaeom}) gives:
\bea
    & & n\delta ^{\mu \nu } \left( {\partial _\mu  \ln H} \right)\left(
    {\partial _\nu  \ln H} \right) + m^2 \kappa ^2
    H^{2\left( {m - n} \right)} \delta ^{\mu \nu } \left( {\partial _\mu
    \ln H} \right)\left( {\partial _\nu  \ln H} \right) \nonumber\\
    & &+ p^2 \left| \omega  \right|^2    H^{\left( {n + 2p} \right)} \delta ^{\mu \nu }
    \left( {\partial _\mu  \ln H} \right)\left( {\partial _\nu  \ln H}
    \right) = 0,
\eea
irrespective of the choice of $f_\mu ^{\,\,\,\,\nu }$ and $l_\mu
^{\,\,\,\,\nu }$. This is clearly satisfied provided that:
\bea
    & &m = n,\nonumber\\ & &p =  - \frac{n}{2} \nonumber\\
    & &n + m^2 \kappa ^2   + p^2 \left| \omega  \right|^2   = 0.\label{mequaln}
\eea

Combining all three conditions together, we get the algebraic
constraint\footnote{Excluding the trivial case $n=0$, which simply
gives the flat Minkowski spacetime with vanishing
hypermultiplets.}:
\be
    1 + n\kappa ^2   + \frac{n}{4}\left| \omega  \right|^2   =    0.
\ee

The $\chi$ field equation (\ref{chieom2}) yields:
\bea
 & &\omega  \left( {n - 2} \right) l^{\mu \nu } \left( {\partial _\mu  \ln H} \right)\left( {\partial _\nu  \ln H} \right) \nonumber\\
  & &+ i2n\omega\kappa    \left( {\varepsilon _{\alpha\beta }  f^{\alpha\mu } l ^{\beta\nu } } \right)\left( {\partial _\mu  \ln H} \right)\left( {\partial _\nu  \ln H} \right) =
  0,\label{chieom3}
\eea
with a similar equation for $\bar\chi$. The conditions for
(\ref{chieom3}) to be satisfied depend on $f_\mu^{\,\,\,\,\nu }$
and $l_\mu^{\,\,\,\,\nu }$ in the following way:

\begin{enumerate}
  \item $f_\mu^{\,\,\,\,\nu }=l_\mu^{\,\,\,\,\nu }=\varepsilon _\mu^{\,\,\,\,\nu }$: Identically satisfied
  \item $f_\mu^{\,\,\,\,\nu }=l_\mu^{\,\,\,\,\nu }=\delta_\mu^\nu $: $\omega\left( {n - 2} \right) = 0$
  \item $f_\mu^{\,\,\,\,\nu }=\delta_\mu^\nu$, $l_\mu^{\,\,\,\,\nu }=\varepsilon_\mu^{\,\,\,\,\nu }$: $\omega\kappa  = 0$
  \item $f_\mu^{\,\,\,\,\nu } = \varepsilon_\mu^{\,\,\,\,\nu } $, $l_\mu^{\,\,\,\,\nu } = \delta_\mu^\nu $: $\omega \left( {n - 2} \right) = \omega \kappa  =0$
\end{enumerate}

Next, the Einstein equation $G_{ab}=8\pi T_{ab}$ is satisfied for
any choice of $f_\mu^{\,\,\,\,\nu }$ and $l_\mu^{\,\,\,\,\nu }$
if:
\be
    B = \frac{n}{4}\left( {1 + \kappa ^2   + \frac{1}{2}\left| \omega  \right|^2  }    \right),
\ee
while $G_{\mu\nu}=8\pi T_{\mu\nu}$ results in:
\bea
 & &\frac{{n^2 }}{2}\left( {\partial _\mu  \ln H} \right)\left( {\partial _\nu  \ln H} \right) - \frac{{n^2 }}{4}\delta _{\mu \nu } \delta ^{\alpha \beta } \left( {\partial _\alpha  \ln H} \right)\left( {\partial _\beta  \ln H} \right) \nonumber\\
 & & + \frac{1}{4}m^2 \kappa ^2  \delta _{\mu \nu } \delta ^{\alpha \beta } \left( {\partial _\alpha  \ln H} \right)\left( {\partial _\beta  \ln H} \right) - \frac{1}{2}p^2 \left| \omega  \right|^2  \delta _{\mu \nu } \delta ^{\alpha \beta } \left( {\partial _\alpha  \ln H} \right)\left( {\partial _\beta  \ln H} \right) \nonumber\\
 & & - \frac{1}{2}m^2 \kappa ^2  f_\mu ^{\,\,\,\,\alpha } f_\nu ^{\,\,\,\,\beta } \left( {\partial _\alpha  \ln H} \right)\left( {\partial _\beta  \ln H} \right) + p^2 \left| \omega  \right|^2  l_\mu ^{\,\,\,\,\alpha } l_\nu ^{\,\,\,\,\beta } \left( {\partial _\alpha  \ln H} \right)\left( {\partial _\beta  \ln H} \right) =
  0.
\eea

This equation yields four algebraic conditions corresponding to the
possibilities:

\begin{enumerate}
    \item $f_\mu^{\,\,\,\,\nu }=l_\mu^{\,\,\,\,\nu}=\varepsilon_\mu^{\,\,\,\,\nu }$: $1 + \kappa ^2   - \frac{1}{2}\left| \omega  \right|^2 =0$
    \item $f_\mu^{\,\,\,\,\nu }=l_\mu^{\,\,\,\,\nu}=\delta_\mu^\nu$: $1 - \kappa ^2   + \frac{1}{2}\left| \omega  \right|^2   = 0$
    \item $f_\mu^{\,\,\,\,\nu }=\delta_\mu^\nu$, $l_\mu^{\,\,\,\,\nu}=\varepsilon_\mu^{\,\,\,\,\nu }$: $1 - \kappa ^2   - \frac{1}{2}\left| \omega  \right|^2 =0$
    \item $f_\mu^{\,\,\,\,\nu } = \varepsilon_\mu^{\,\,\,\,\nu } $, $l_\mu^{\,\,\,\,\nu} = \delta_\mu^\nu $: $1 + \kappa ^2   + \frac{1}{2}\left| \omega  \right|^2  =0$
\end{enumerate}

We now solve all the algebraic conditions under the four possible
combinations of $f_\mu^{\,\,\,\,\nu }$ and $l_\mu^{\,\,\,\,\nu }$.
We choose the complex representation $\omega=b e^{i\varphi}$,
where $b$ is a real constant and $\varphi$ is an arbitrary phase
and demand that $\kappa=\bar\kappa$ to ensure the reality of $F$.
We find \emph{exactly} two possibilities:
\begin{description}
    \item[2-brane with constant axions] \bea
            ds^2  &=& \eta _{ab} dx^a dx^b  + e^{-\sigma}\delta _{\mu \nu } dx^\mu  dx^\nu  ,\nonumber\\ \sigma  &=&   - \ln H,\quad\quad d\chi  = d\bar \chi  =
            0 \nonumber\\
            F &=&  \pm  \left( {\partial _\mu  H^{ - 1} } \right)dt \wedge dx^1  \wedge dx^2  \wedge dx^\mu,\nonumber\\
            \to A &=&  \pm \frac{1}{H}dt \wedge dx^1  \wedge dx^2,\quad \quad \quad \left( {\chi ,\bar \chi } \right) = {\rm constants}.\label{C}
        \eea
    \item[2-brane with constant 3-form field] \bea
             ds^2  &=& \eta _{ab} dx^a dx^b  + e^{-2\sigma} \delta _{\mu \nu } dx^\mu  dx^\nu  ,\nonumber\\ \sigma  &=&   - 2\ln H,\quad \quad F = 0,\quad\quad A= {\rm constant} \nonumber\\
             d\chi &=&  \pm \sqrt 2   \varepsilon _\mu ^{\,\,\,\,\nu } \left( {\partial _\nu  H} \right)e^{ + i\varphi }dx^\mu,\quad \quad \quad
             d\bar\chi =  \pm \sqrt 2   \varepsilon _\mu ^{\,\,\,\,\nu } \left( {\partial _\nu  H} \right)e^{ - i\varphi
             }dx^\mu,\nonumber\\
             \to \chi &=&  \pm \sqrt 2 qe^{i\varphi } \theta,\quad \quad \quad
             \bar\chi =  \pm \sqrt 2 qe^{-i\varphi } \theta.\label{D}
        \eea
\end{description}

It is straightforward to verify that the conditions $\delta
\psi=\delta\xi=0$ are satisfied if the SUSY spinors
$\left(\epsilon_1,\epsilon_2\right)$ are constant in both cases.
Solutions (\ref{C}) and (\ref{D}) are exactly the ones we found in
reference \cite{Emam:2005bh} as the result of the dimensional
reduction of wrapped M-branes and are the \emph{only} 2-brane
possibilities satisfying the ans\"{a}tze (\ref{ansatze1},
\ref{sigmageneral}, \ref{ansatze2}). We have discussed the M-brane
connection in more detail in \cite{Emam:2005bh} and
\cite{Emam:2006sr} (as well as \cite{Emam:2004nc}). In the first
reference in particular, it was explicitly shown that the
dimensional reduction of a single M2-brane down to five dimensions
over a rigid Calabi-Yau manifold $(\htwo=0)$ gives exactly the
solution (\ref{C}), while the dimensional reduction of a certain
M5-brane configuration yields the solution (\ref{D})\footnote{In
\cite{Emam:2005bh}, however, the choice $\varphi {\rm = }{\pi
\mathord{\left/ {\vphantom {\pi {\rm 2}}} \right.
\kern-\nulldelimiterspace} {\rm 2}}$ was used.}. It is clear from
the calculations in this section that these are the only possible
2-brane solutions if one insists on applying the dependence on $H$
to all the UH fields. To find a solution with non-vanishing fields
we will have to relax this condition, which we will do in the next
section.

Finally, we note that if the conditions $\kappa=\bar \kappa$ and
$b=\bar b$ are relaxed then there exist two more solutions:
\bea
                ds^2  &=& \eta _{MN} dx^M dx^N ,\nonumber\\ \sigma  &=&  \ln H, \quad  d\chi  = d\bar \chi  =0 \nonumber\\
                F &=&  \pm i  \varepsilon _\mu ^{\,\,\,\,\nu } \left( {\partial _\nu  H} \right)dt \wedge dx^1  \wedge dx^2  \wedge dx^\mu,\nonumber\\
                \to A &=& \pm iq\theta\, dt \wedge dx^1  \wedge dx^2,  \quad \quad \quad \left( {\chi ,\bar \chi } \right) = {\rm constants},\label{A}
\eea
and
\bea
                ds^2  &=& \eta _{MN} dx^M dx^N ,\nonumber\\ \sigma  &=&   2\ln H,\quad \quad F = 0 \nonumber\\
                d\chi &=&  \pm i\sqrt 2  \left( {\partial _\mu  H^{ - 1} } \right)e^{ + i\varphi }dx^\mu,\quad \quad \quad
                d\bar\chi =  \pm i\sqrt 2  \left( {\partial _\mu  H^{ - 1} } \right)e^{ - i\varphi
                }dx^\mu\nonumber\\
                \to \chi &=&   \pm i\frac{{\sqrt 2 }}{H}e^{i\varphi },\quad \quad \quad
                \bar\chi =   \pm i\frac{{\sqrt 2 }}{H}e^{-i\varphi }, \quad\quad A = {\rm constant}.\label{B}
\eea

These are clearly not 2-branes but rather Minkowski space
representations of Euclidean solutions of the theory, \emph{i.e.}
instantons; as can be verified by performing a Wick rotation $t
\to ix^0 $ on them and allowing the function $H$ to be harmonic in
all five dimensional Euclidean space (which also restores the
reality of $F$ in the first result). Such UH-coupled instantons
have been extensively studied in the literature, particularly for
$D=4$ (see for example
\cite{Becker:1995kb,Becker:1999pb,Davidse:2003ww,Davidse:2004gg}).
The first solution (\ref{A}) is the result of the dimensional
reduction of Euclidean M5-branes over a rigid $\M$ and (\ref{B})
is the dimensional reduction of Euclidean M2-branes. Solutions
(\ref{A}) and (\ref{B}) are then just the Minkowski space
representations of these instantons and are magnetically dual to
the 2-branes (\ref{C}) and (\ref{D}). The $D=4$ instantons of
\cite{Becker:1995kb} are similar and should arise as the
dimensional reduction of (\ref{A}) and (\ref{B}) over $S^1$.

\section{The 2-brane with full UH fields}

In the previous section it was shown that assuming the ansatz (\ref{ansatze2}) fails to deliver a
solution with non-vanishing UH fields. We now relax
(\ref{ansatze2}) while keeping the analysis of
\S\ref{HarmonicAnsatzMain}. Based on the form of the solutions
(\ref{C}), (\ref{D}), (\ref{A}) and (\ref{B}), we expect that the
choice $f_\mu^{\,\,\,\,\nu }=\delta_\mu^\nu$ will lead to a
2-brane configuration generalizing (\ref{C}), while
$f_\mu^{\,\,\,\,\nu } = \varepsilon_\mu^{\,\,\,\,\nu } $ would
result in an instanton solution\footnote{This, in fact, leads to
the solution found in \cite{Gutperle:2000sb}, which reduces to
(\ref{A}) for vanishing axions.}. We then set
\bea
    ds^2  &=& \eta _{ab} dx^a dx^b  + e^{-\sigma}\delta _{\mu \nu } dx^\mu  dx^\nu  ,\quad \quad \quad \sigma  =   - \ln H \nonumber\\
            F &=&  \pm  \left( {\partial _\mu  H^{ - 1} } \right)dt \wedge dx^1  \wedge dx^2  \wedge
            dx^\mu,
\eea
and are only left with finding the form of the non-constant
axions. First, we consider the Einstein equations $G_{ab}=8\pi
T_{ab}$ and $G_{\mu\nu}=8\pi T_{\mu\nu}$, both of which lead to
\be
        d\sigma  \wedge \star d\sigma  + e^{ - 2\sigma } F \wedge \star F + 2 e^\sigma  d\chi  \wedge \star d\bar \chi  =
        0,\label{Einstein3}
\ee
yielding
\bea
    d\chi  \wedge \star d\bar \chi  =  0.\label{Cauchy1}
\eea

We also note that the dilaton equation (\ref{sigmaeom}) leads to (\ref{Cauchy1}) as well. It can be shown that it is impossible to satisfy (\ref{Cauchy1}) simultaneously with (\ref{dchiwedgedchibar}), which guarantees (\ref{Feom}), if one insists that the axions $(\chi, \bar\chi)$ are complex conjugate to each other. However, this problem is solved if one assumes that they are \emph{split}-complex conjugate. In other words:
\bea
    \chi  &=& \chi _1  + j\chi _2  \nonumber\\
    \bar \chi  &=& \chi _1  - j\chi _2,
\eea
where $\left(\chi _1, \chi _2\right)$ are functions in the transverse plane and the ``imaginary'' number $j$ is defined by $j^2=+1$ and is \emph{not} equal to $\pm1$. Split-complex numbers\footnote{Also known as `para-complex numbers', `real tessarines', `algebraic motors', `hyperbolic complex numbers', `double numbers', `perplex numbers', `Lorentz numbers', and several others.} are a generalization of ordinary complex numbers satisfying the `hyperbolic' scalar product:
\be
    \left| \chi  \right|^2  = \chi _1^2  - \chi _2^2.
\ee

In contrast to the complex numbers, which form a field, the split-complex numbers form a ring. They have the interesting property, absent from the complex numbers, of containing non-trivial idempotents (other than 0 and 1), where an idempotent $Z$ is defined by $Z^2=Z$. This property can be used to define the so-called diagonal, or null, basis:
\bea
    e &=& \frac{1}{2}\left( {1 + j} \right) \nonumber\\
    \bar e &=& \frac{1}{2}\left( {1 - j} \right),
\eea
such that any split-complex quantity, such as our axion fields, can be written in the form:
\bea
    \chi  &=& \left( {\chi _1  + \chi _2 } \right)e + \left( {\chi _1  - \chi _2 } \right)\bar e \\
    \bar \chi  &=& \left( {\chi _1  - \chi _2 } \right)e + \left( {\chi _1  + \chi _2 } \right)\bar e,
\eea
where $e$ is idempotent as well as null, \emph{i.e.} $\left| e \right|^2  = e\bar e = 0$. One can also demonstrate the interesting property (which we will use in the next section):
\be
    \frac{1}{e} = \frac{1}{{\bar e}} = 1.\label{interestingproperty}
\ee

Equations (\ref{Cauchy1}) and (\ref{dchiwedgedchibar}) further require that $\chi_1=\chi_2$, \emph{i.e.}
\bea
    \chi  &=& 2e\chi _1  \nonumber\\
    \bar \chi  &=& 2\bar e\chi _1 ,
\eea
which means that $\left(\chi, \bar\chi\right)$ are themselves null split-complex fields. It is this last fact that allows for (\ref{Cauchy1}) and (\ref{dchiwedgedchibar}) to be simultaneously satisfied. It is interesting to note that split-complex structures have been shown to naturally arise in the context of hypermultiplet couplings obeying Euclidean supersymmetry (see \emph{e.g.} \cite{Cortes:2005uq}
 and the references within). The connection between this and the explicit solution found here might be a worthwhile problem to pursue.

We are now left with finding the explicit form of the real function $\chi _1 \left( {r,\theta } \right)$. This is straightforwardly found by looking at the $\chi$ field equation (\ref{chieom2}), which gives
\be
    \frac{1}{r}\frac{{\partial \,\,}}{{\partial r}}\left( {r\frac{{\partial \chi_1 }}{{\partial r}}} \right) + \frac{1}{{r^2 }}\left( {\frac{{\partial ^2 \chi_1 }}{{\partial \theta ^2 }}} \right) = \frac{q}{Hr}\left[ {\left( {\frac{{\partial \chi_1 }}{{\partial r}}} \right) + \frac{i}{{r}}\left( {\frac{{\partial \chi_1 }}{{\partial \theta }}} \right)\,} \right].\label{chieom4}
\ee

This clearly implies that
\be
    \frac{{\partial \chi_1 }}{{\partial \theta}} = 0.
\ee

The remaining radial equation is then readily integrated to give:
\bea
    \chi  &=& \chi_0 H^2 e \nonumber\\
    \bar \chi  &=& \chi_0 H^2 \bar e,
\eea
where the factor of 2 was absorbed in the arbitrary integration constant $\chi_0$. Note that the axion fields are asymptomatically well-behaved since $\left(d\chi, d\bar\chi\right)  \propto {H \mathord{\left/
 {\vphantom {H r}} \right. \kern-\nulldelimiterspace} r}$ ; slowly converging to a finite value.

Finally, we look at the SUSY variations. Using (\ref{hyperSUSY}),
the condition $\delta \xi=0$ can be written in the matrix form
\be
    \left[ {\begin{array}{*{20}c}
   {\frac{1}{2}\left[ {\left( {\partial _\mu \sigma } \right) - ie^{ - \sigma } \left( {\star F} \right)_\mu } \right]\Gamma ^\mu} & {\frac{{e^{\frac{\sigma }{2}} }}{{\sqrt 2 }}\left( {\partial _\mu \chi } \right)\Gamma ^\mu }  \\
   {- \frac{{e^{\frac{\sigma }{2}} }}{{\sqrt 2 }}\left( {\partial_\nu\bar \chi } \right)\Gamma ^\nu} & {\frac{1}{2}\left[ {\left( {\partial _\nu \sigma } \right) + ie^{ -\sigma } \left( {\star F} \right)_\nu } \right]\Gamma ^\nu }  \\
    \end{array}} \right]\left( {\begin{array}{*{20}c}
   {\epsilon _1}  \\
   { \epsilon _2}  \\
    \end{array}} \right) = 0,
\ee
which is true if and only if the determinant of the given matrix
vanishes:
\be
    d\sigma  \wedge \star d\sigma  + e^{ - 2\sigma } F \wedge \star F + 2 e^\sigma  d\chi  \wedge \star d\bar \chi  =
    0.
\ee

This is exactly equation (\ref{Einstein3}) which we already know
is satisfied. The condition $\delta \xi=0$ can further be used to
uncouple the two $\delta\psi=0$ equations, leading to first order
differential equations in the spinors:
\bea
    D\epsilon _1  + \frac{1}{2}d\sigma \epsilon _1  - \frac{i}{4}e^{ - \sigma } \left( {\star F} \right)\epsilon _1  &=& 0 \nonumber\\
    D\epsilon _2  + \frac{1}{2}d\sigma \epsilon _2  + \frac{i}{4}e^{ - \sigma } \left( {\star F} \right)\epsilon _2  &=&
    0,
\eea
resulting in
\be
    \left( {\frac{{d \epsilon _s }}{{d r}}} \right) = \frac{q}{{2Hr}}\epsilon _s \,\,\,\,\,\,\,\,\,\,\,\,\,\,\,\,\,\,\,s =
    1,2,
\ee
which is straightforwardly solved by $\epsilon _s  = H^{\frac{1}{2}} \hat \epsilon _s$, where $\left(\hat\epsilon_1,\hat\epsilon_2\right)$ are constant
spinors.

To summarize, the complete solution representing the coupling of a
2-brane to the full set of non-vanishing universal hypermultiplet
fields in $D=5$ is:
\bea
    ds^2  &=& \eta _{ab} dx^a dx^b  + e^{-\sigma}\delta _{\mu \nu } dx^\mu  dx^\nu  ,\quad \quad \quad \sigma  =   - \ln H \nonumber\\
    A &=&  \pm \frac{1}{H}dt \wedge dx^1  \wedge dx^2,\nonumber\\
    \chi  &=& \chi_0 H^2 e \quad\quad\quad\quad{\rm where}\quad e = \frac{1}{2}\left( {1 + j} \right), \quad
    \bar e = \frac{1}{2}\left( {1 - j} \right),\nonumber\\
    \bar \chi  &=& \chi_0 H^2 \bar e, \quad\quad\quad\quad{\rm and}\quad j^2=+1, \quad j\ne \pm 1.\label{Final Solution}
\eea

This clearly reduces to (\ref{C}) if $\chi _0=0$ as required.

At this point, a natural question to ask is whether or not one can
find another solution with non-vanishing UH fields that reduces to
(\ref{D}) if $F=0$. One can show that this, in fact, is not
possible. The reason is that if one assumes such a solution
existed then equation (\ref{Einstein3}) necessarily leads to
\be
    F \wedge \star F = 0\quad \to\quad \left( {\frac{{\partial A}}{{\partial r}}} \right)^2  + \frac{1}{{r^2 }}\left( {\frac{{\partial A}}{{\partial \theta }}}
    \right)^2=0,\label{NoRealA}
\ee
implying that $A$ is not real. This is a contradiction and
one concludes that a more general solution, exactly reducing to
(\ref{D}) as a special case, does not exist. However, it should be
clear that relaxing the condition of the reality of $F$ and
proceeding from (\ref{NoRealA}) one should arrive at an instanton
solution that reduces to (\ref{B}) for vanishing $F$. As before,
the reality of $F$ would be restored in Euclidean space.

\section{The M-branes connection revisited}

In order to emphasize the geometric meaning of the results
discussed in this paper, we further analyze their interpretation
as the dimensional reduction of M-brane configurations. The
M-brane charges, excited by these reductions, are calculable via
(\ref{Charges}). Note however, that an interesting property of the
2-branes (\ref{C}), (\ref{D}) and (\ref{Final Solution}) is that
they are asymptotically non-flat! This phenomenon is known to
arise for $p$-branes where $p=D-3,D-2$ and such solutions have
been dubbed `high branes' \cite{Bergshoeff:1996uv}. As such, it
does not make much sense to talk about the `total' charge of the
brane, since the integrals of (\ref{Charges}) would necessarily
diverge if carried over the entire transverse space. In what
follows then, we calculate the M-brane charges using specific
choices of integration constants, with the sole purpose of
relating $\left(\mathcal{Q}_2, \mathcal{Q}_5\right)$ to the
dilaton charge $q$ in the simplest possible way. We will use
non-script $\left(Q_2, Q_5\right)$ to denote the charges with
these choices of constants.

As discussed in \cite{Emam:2005bh}, (\ref{C}) is the result of
reducing a single M2-brane to $D=5$ using the scheme
\be
    \begin{array}{*{20}c}
   t & 1 & 2 & \vline &  3  &  4  & \vline &  \bullet  &  \bullet  &  \bullet  &  \bullet  &  \bullet  &  \bullet   \\
    \end{array}
\ee
where $\left(t,1,2\right)$ represent the world-volume directions
of the brane, $\left(3,4\right)$ the transverse directions and the
six dimensions denoted by $\bullet$ represent those wrapped
completely over a $T^6$. Calculating the M2-brane charge via
(\ref{Charges}) leads to the simple relation
\be
    Q_2=\pm q,
\ee
and of course $Q_5$ vanishes. This leads to
\bea
    ds^2  &=& \eta _{ab} dx^a dx^b  + e^{-\sigma}\delta _{\mu \nu } dx^\mu  dx^\nu  ,\nonumber\\ \sigma  &=&   - \ln \left(1+q \ln r\right) \nonumber\\
    A &=&   \frac{1}{{1 + Q_2 \ln r}} dt \wedge dx^1  \wedge dx^2,\quad\quad \left(\chi,\bar\chi\right)={\rm constants}.\label{C1}
\eea

Solution (\ref{D}), on the other hand, is slightly more complex
since it arises from the dimensional reduction of a system of 4
intersecting M5-branes as follows
\be
    \begin{array}{l}
 \begin{array}{*{20}c}
   t & 1 & 2 & \vline & 3 & 4 & \vline & 5 & 6 & 7 &  \bullet  &  \bullet  &  \bullet   \\
\end{array} \\
 \begin{array}{*{20}c}
   t & 1 & 2 & \vline & 3 & 4 & \vline & 5 &  \bullet  &  \bullet  &  \bullet  & 9 & {10}  \\
\end{array} \\
 \begin{array}{*{20}c}
   t & 1 & 2 & \vline & 3 & 4 & \vline &  \bullet  & 6 &  \bullet  & 8 &  \bullet  & {10}  \\
\end{array} \\
 \begin{array}{*{20}c}
   t & 1 & 2 & \vline & 3 & 4 & \vline &  \bullet  &  \bullet  & 7 & 8 & 9 &  \bullet   \\
\end{array} \\
    \end{array}\label{M5Array}
\ee

Note that each two M5-branes intersect on a 1-brane and that three
directions of each M5-brane end up completely wrapped over
3-cycles of the Calabi-Yau, which are taken to be special
Lagrangian. The $Q_2$ charge vanishes and we find:
\bea
    Q_5  &=&  \pm 2\pi \sqrt 2 qe^{i\varphi } ,\nonumber\\ \bar Q_5  &=&  \pm 2\pi \sqrt 2 qe^{ - i\varphi
    },
\eea
leading to
\bea
    ds^2  &=& \eta _{ab} dx^a dx^b  + e^{-2\sigma} \delta _{\mu \nu } dx^\mu  dx^\nu  ,\nonumber\\ \sigma  &=&   - 2\ln (1+q\ln r), \nonumber\\
    \chi &=&  \frac{{Q_5 }}{{2\pi }}\theta,\nonumber\\
    \bar\chi &=&  \frac{{\bar Q_5 }}{{2\pi }}\theta, \quad\quad A={\rm constant}.\label{D1}
\eea

The fact that one cannot find a more general solution with
non-vanishing charges that reduces to (\ref{D1}) can be explained
in the following way: Such a solution, if it existed, would have
the $D=11$ interpretation as the intersection of four M5-branes
and at least one wrapped M2-brane. Chart (\ref{M5Array}) implies,
however, that there is no room left for M2-branes. If the $D=11$
parent configuration does not exist, then the $D=5$ offspring
wouldn't either, as we found in the previous section.

Finally, we look at (\ref{Final Solution}). The M-brane charges
are found to be:
\bea
    Q_2  &=&  \pm q  \nonumber\\
    Q_5  &=& \pm 2q\chi _0 e\left( {1 - i\pi } \right)\nonumber\\ \bar Q_5  &=& \pm 2q\chi _0 \bar e\left( {1 + i\pi } \right).
\eea

Note that the M5-brane charges $\left(Q_5, \bar Q_5\right)$ are both complex \emph{and} null split-complex. Re-writing $\chi_0$ in terms of the M-brane charges and using (\ref{interestingproperty}), (\ref{Final Solution}) becomes:
\bea
    ds^2  &=& \eta _{ab} dx^a dx^b  + e^{-\sigma}\delta _{\mu \nu } dx^\mu  dx^\nu  ,\quad \quad \quad \sigma  =   - \ln\left( {1 + q\ln r} \right) \nonumber\\
    A &=&  \pm \frac{1}{\left( {1 + Q_2 \ln r} \right)}dt \wedge dx^1  \wedge dx^2,\nonumber\\
    \chi  &=&  \pm \frac{{Q_5 }}{{Q_2 }}\frac{{e\left( {1 + i\pi } \right)}}{{2\left( {1 + \pi ^2 } \right)}}\left( {1 + Q_2 \ln r} \right)^2  \nonumber\\
    \bar \chi  &=&  \pm \frac{{\bar Q_5 }}{{Q_2 }}\frac{{\bar e\left( {1 - i\pi } \right)}}{{2\left( {1 + \pi ^2 } \right)}}\left( {1 + Q_2 \ln r} \right)^2 ,\nonumber\\
    & &{\rm where}\quad e = \frac{1}{2}\left( {1 + j} \right), \quad
    \bar e = \frac{1}{2}\left( {1 - j} \right)  \nonumber\\
    & &{\rm and}\quad j^2=+1, \quad j\ne \pm 1.\label{Final Solution2}
\eea

Solution (\ref{Final Solution}/\ref{Final Solution2}) then represents the dimensional
reduction of some configuration of wrapped M2- and M5-branes.
Finding this configuration is an interesting problem in its own
right. It might be, however, that it is already partially known.
In \cite{Izquierdo:1995ms}, a solution representing the coupling
of (unwrapped) M2- and M5-branes was found (with further
discussion in \cite{Green:1996vh}). In their analysis, the $D=11$
metric and 4-form field $\mathcal{F}$ are dependent on an
arbitrary phase angle $\alpha$ such that the choice $\cos \alpha =
0$ reduces the solution to a pure M2-brane whereas $\sin \alpha  =
0$ gives the ordinary M5-brane. We suspect then that there exists
an eleven dimensional solution that represents the wrapping of
this dyonic state over special Lagrangian cycles of $\M$ such
that, upon dimensional reduction, it yields (\ref{Final
Solution}). This, however, is a question for future pondering.

\section{ Conclusion}

We presented a detailed analysis of the coupling of supersymmetric
BPS 2-brane sources to the universal hypermultiplet fields of five
dimensional $\N=2$ SUGRA. These solutions can be thought of as the
dimensional reduction of certain M-brane configurations over
special Lagrangian cycles of a rigid Calabi-Yau 3-fold. Two of the
solutions discussed carry only one of the two possible M-brane
charges and were shown to be the only solutions satisfying both
(Poincar\'{e})$_3 \times SO\left( 2 \right)$ as well as the
dependence on a purely radial harmonic function (barring instanton
solutions of the Euclidean form of the theory). We then
generalized one of these solutions to one representing the
dimensional reduction of a coupled M2/M5 brane configuration and
explicitly calculated the M-brane charges.

There are several possible paths that one can tread from here. For
example, finding the eleven dimensional dyonic M2/M5 brane
configuration that dimensionally reduces to (\ref{Final Solution}/\ref{Final Solution2})
would be interesting. As mentioned, a starting point could be the
solution of \cite{Izquierdo:1995ms}. One may also generalize our
result to include an arbitrary number of hypermultiplet fields
(\textit{i.e.} $\htwo\ne 0$) in the spirit of \cite{Emam:2005bh}
and \cite{Gutperle:2000ve}, and/or find its eleven dimensional
counterpart. Clearly these possibilities are intertwined and one
may easily lead to the other. An obvious obstacle would be the
fact that explicit forms of compact non-rigid Calabi-Yau manifolds
do not yet exist. In other words any $\htwo\ne 0$ configuration
will necessarily depend on unknown functions representing the
complex structure moduli and other metric-dependent properties of
$\M$. Despite this, one can still deduce a variety of conclusions
from the general form of the differential equations governing the
unknown functions, as well as further understand Calabi-Yau
compactifications (see \textit{e.g.} \cite{Kastor:2003jy} and
\cite{Emam:2006sr}). Furthermore, constructing wrapped M-branes
may require invoking the theory of generalized calibrations; an
approach that is interesting in its own right (\textit{e.g.}
\cite{Cho:2000hg}).

Even without reference to higher dimensional sources, one can
still find open questions to tackle for hypermultiplets-coupled
branes. For instance, it would be interesting to further explore
the properties of $\N=2$ branes as high branes. Also, aside from the appearance of split-complex structures in Euclidean supersymmetry \cite{Cortes:2005uq}, the use of split-complex quantities to construct an explicit supergravity solution is, to this author's knowledge, unprecedented. One could further explore the origins of this, its connection to Euclidean supersymmetry, as well as its physical implications, if any. Another
possibility, one more in tune with our earlier comment that
hypermultiplets-related research is lacking in the literature,
would be to write down general ans\"{a}tze for five dimensional
$p$-branes satisfying (Poincar\'{e})$_{p+1} \times SO\left( D-p-1
\right)$ invariance and study the general properties of such
metrics and how they can couple to the UH fields or, more
ambitiously, to the full hypermultiplets sector. Some work in this
direction already exists \cite{Ketov:2001gq},
\cite{Bellorin:2006yr}. This is particularly interesting in the
context of classifying $p$-branes in supergravity. In the future,
we plan to continue explorations in at least some of the above
possibilities.

\section*{Acknowledgments}

Thanks are due to my friend and colleague Mohamed Anber for useful
discussions and advice.

\pagebreak

\end{document}